\documentclass{ws-mpla}

\begin{document}

\markboth{Y. Ping et al.} {$5D$ Solutions to $\Lambda$CDM Universe
Derived from Global Brane Model}

%
\catchline{}{}{}{}{}
%
\title{$5D$ Solutions to $\Lambda$CDM Universe
Derived from Global Brane Model}

\author{Yongli Ping, Lixin Xu, Baorong Chang, Molin Liu and Hongya Liu}

\address{School of Physics and Optoelectronic Technology,\\ Dalian
University of Technology, Dalian, Liaoning 116024,
P.R.China\\ylping@student.dlut.edu.cn}

\maketitle

\pub{Received }{Revised }

\begin{abstract}
An exact solution of brane universe is studied and the result
indicates that Friedmann equations on the brane are modified with an
extra term. This term can play the role of dark energy and make the
universe accelerate. In order to derive the $\Lambda$CDM Universe
from this global brane model, the new solutions are obtained to
describe the $5D$ manifold.
\end{abstract}

 \ccode{PACS numbers:
04.50.+h, 98.80.-k, 02.40.-k}

\section{Introduction}
Recent observations indicate that our universe is undergoing
accelerated expansion\cite{Riess,Perlmutter} and dominated by a
negative pressure component dubbed dark energy. Obviously, a natural
candidate to dark energy is a cosmological constant with equation of
state $w_{\Lambda}=-1$. Einstein (1917) introduced the cosmological
constant $\Lambda$, because he believed that the universe is
static.\cite{Einstein} However, Friedmann (1922) discovered an
expanding solution to the Einstein field equations in the absence of
$\Lambda$ and Hubble (1929) found the universe was
expanding.\cite{Friedmann,Hubble} Soon after, Einstein discarded the
cosmological constant and admitted his greatest blunder. Although
abandoned by Einstein, the cosmological constant staged several
come-backs. It was soon realized that, since the static Einstein
universe is unstable to small perturbations, one could construct
expanding universe models which had a quasi-static origin in the
past, thus ameliorating the initial singularity which plagues
expanding FRW models. Theoretical interest in $\Lambda$ remained on
the increase during the 1970¡¯s and early 1980¡¯s with the
construction of inflationary models, in which matter (in the form of
a false vacuum, as vacuum polarization or as a minimally coupled
scalar-field) behaved precisely like a weakly time-dependent
$\Lambda$-term. The cosmological constant makes an important
appearance in models with spontaneous symmetry
breaking.\cite{Weinberg} The current interest in $\Lambda$ stems
mainly comes from observations of Type Ia high redshift supernovae
which indicate that the universe is accelerating expansion fueled
perhaps by a small cosmological
$\Lambda$-term.\cite{Riess,Perlmutter} The review about the
cosmological constant can be seen.\cite{Weinberg,Varun Sahni,Varun
Sahni1,Varun Sahni2}

It is proposed that our universe is a 3-brane embedded in a
higher-dimensional
space.\cite{Arkani-Hamed,Arkani-Hamed2,Arkani-Hamed3,Horava,Witten,Witten2,Randall,Randall2}
In brane-world model, gravity can freely propagate in all
dimensions, while standard matter particles and forces are confined
on the 3-brane. A five-dimensional ($5D$) cosmological model and
derived Friedmann equations on the branes are considered by
Binetruy, Deffayet and Langlois (BDL),\cite{BDL} for a recent
review, it can be seen.\cite{Marttens,Philippe Brax} Brane-world
models of dark energy are studied\cite{Sahni2} and accelerating
universe comes from gravity leaking to extra dimension in DGP
brane.\cite{Deffayet}

In this paper, we derive $\Lambda$CDM universe from global brane
model with a Ricci-flat bulk characterized by a class of exact
solutions. The solutions were firstly presented by Liu and Mashhoon
and restudied latter by Liu and Wesson.\cite{L-M,L-W} And these
solutions are algebraically rich because they contain two arbitrary
functions of time $t$. The solutions are utilized in
cosmology\cite{X-W,W-X,XU,CHANG,L-L,X-Z,ZHANG,PING,Meng} and are
relate to the brane model.\cite{Seahra,J. Ponce de Leon,Liu} In
order to induce $\Lambda$CDM universe from global brane model, more
exact solutions of the $5D$ bulk are obtained.
\section{Friedmann equations in global brane universes}
A class of $5D$ Ricci-flat cosmological solution reads\cite{L-M}
\begin{equation}
dS^{2}=B^{2}dt^{2}-A^{2}\left( \frac{dr^{2}}{1-kr^{2}}+r^{2}d\Omega
^{2}\right) -dy^{2}, \label{line element}
\end{equation}%
\begin{equation}
A^{2}=\left( \mu ^{2}+k\right) y^{2}+2{\nu }y+\frac{\nu ^{2}+K}{\mu
^{2}+k} ,  \label{A}
\end{equation}%
\begin{equation}
B=\frac{1}{\mu }\frac{\partial A}{\partial t}\equiv
\frac{\dot{A}}{\mu },  \label{B}
\end{equation}%
where $d\Omega ^{2}=d\theta ^{2}+\sin^2 \theta d\psi ^{2}$; $\mu
=\mu \left( t\right) $ and $\nu =\nu \left( t\right) $ are two
arbitrary functions of time $t$; $k$ is the 3D curvature index
($k=\pm 1,0$), and $K$ is a constant. Because the $5D$ manifold
(\ref{line element})-(\ref{B}) is Ricci-flat, we have $I_{1}\equiv
R=0$, $I_{2}\equiv R^{AB}R_{AB}=0$, and%
\begin{equation}
I_{3}\equiv R^{ABCD}R_{ABCD}=\frac{72K^{2}}{A^{8}} ,  \label{I-3}
\end{equation}%
so $K$ is related to the $5D$ curvature. They are used as the bulk
solutions of the BDL-type brane model. To obtain brane models for
using the $Z_2$ reflection symmetry on $A$ and $B$, they are set
as\cite{Liu}
\begin{eqnarray}
A^{2}&=&\left( \mu ^{2}+k\right) y^{2}-2{\nu}\mid{y}\mid+\frac{\nu
^{2}+K}{\mu ^{2}+k} ,  \nonumber\\
B&=&\frac{1}{\mu }\frac{\partial A}{\partial t}\equiv
\frac{\dot{A}}{\mu }.  \label{A}
\end{eqnarray}%

Then the corresponding $5D$ bulk Einstein equations are taken as
\begin{eqnarray}
G_{AB}&=&\kappa^2_{(5)}T_{AB}, \nonumber\\
T_B^A&=&\delta(y)diag(\rho_1,-p_1,-p_1,-p_1,0)  \nonumber \\
&\ &+\delta(y-y_2)diag(\rho_2,-p_2,-p_2,-p_2,0)\label{G}
\end{eqnarray}
where the first brane is at $y=y_1=0$ and the second is at
$y=y_2>0$. In the bulk $T_{AB}=0$ and $G_{AB}=0$, Eq.(\ref{G}) are
satisfied by (\ref{A}). On the branes, Liu had solved Eq.(\ref{G})
in Ref.~\refcite{Liu}. We adopt the result at $y=y_1=0$ and
$y=y_2>0$ as follows:
\begin{eqnarray}
\kappa^2_{(5)}{\rho}_1&=&\frac{6\nu}{A_1^2}, \label{rho1}\\
\kappa^2_{(5)}p_1&=&-\frac{2}{\dot{A_1}}\frac{\partial}{\partial{t}}(\frac{\nu}{A_1})-\frac{4\nu}{A^2_1}
\label{p1},
\end{eqnarray}
and
\begin{eqnarray}
\kappa^2_{(5)}{\rho}_2&=&\frac{6}{A_2}(\frac{\mu^2+k}{A_2}y_2-\frac{\nu}{A_2}), \label{rho2}\\
\kappa^2_{(5)}p_2&=&-\frac{2}{\dot{A_2}}\frac{\partial}{\partial{t}}(\frac{\mu^2+k}{A_2}y_2-\frac{\nu}{A_2}) \nonumber\\
&\ &-\frac{4}{A_2}(\frac{\mu^2+k}{A_2}y_2-\frac{\nu}{A_2})
\label{p2},
\end{eqnarray}
where, $A_1$ is the scale factor on $y=y_1=0$ brane and $A_2$ is the
scale factor on $y=y_2>0$ brane.

Now, we consider the universe on the second brane, i.e. $y=y_2>0$.
From the $5D$ metric (\ref{line element}), the Hubble and
deceleration parameters on the $y=y_2$ brane can be defined as
\begin{eqnarray}
H_2(t,y)&\equiv&\frac{1}{B_2}\frac{\dot{A_2}}{A_2}=\frac{\mu}{A_2},
\label{H}\\
q_2(t,y)&=&-\frac{A_2\dot{\mu}}{\mu\dot{A_2}}.\label{q_2}
\end{eqnarray}
Substituting Eq.(\ref{H}) into Eq.(\ref{rho2}) to eliminate $\mu^2$,
Eq. (\ref{rho2}) can be rewritten into a new form as
\begin{equation}
H^2_2+\frac{k}{A^2_2}=\frac{\kappa^2_{(5)}}{6y_2}({\rho}_2+\frac{6}{\kappa^2_{(5)}}\frac{\nu}{A^2_2}).
\label{friedman e}
\end{equation}
We define $\rho_x=\frac{6}{\kappa^2_{(5)}}\frac{\nu}{A^2_2}$ and it
can play the role of dark energy. Then from the Eq.(\ref{p2}), we
have
\begin{equation}
\frac{2\mu\dot{\mu}}{A_2\dot{A_2}}+\frac{\mu^2+k}{A^2_2}
=-\frac{\kappa^2_{(5)}}{2y_2}\left(p_2-p_x\right),\label{f2}
\end{equation}
where
$p_x=-\frac{2}{\kappa^2_{(5)}}(\frac{\dot{\nu}}{A_2\dot{A}_2}+\frac{\nu}{A^2_2})$.
Meanwhile, the conservation law $T^B_{A;B}=0$ gives
\begin{equation}
\dot{\rho_2}+3(\rho_2+p_2)\frac{\dot{A_2}}{A_2}=0.
\label{conservation}
\end{equation}
From Eqs. (\ref{friedman e}) and (\ref{f2}), it can be seen that the
extra term which will be treated as dark energy have been induced on
the brane.\cite{PING2} By assuming that only dark matter is
contained on the brane, we have $p_2=0$ and
$\rho_2=\rho_{20}A_{20}^3A_2^3$. Then, from Eq. (\ref{friedman e})
and Eq. (\ref{f2}), for $k=0$, EOS of dark energy, dimensionless
density parameters and deceleration parameters $q_2$ with $A_{20}=1$
and $\nu_0=1$ can be obtained
\begin{eqnarray}
w_x&=&\frac{p_x}{\rho_x}=-\frac{1}{3}(\frac{A_2\dot{\nu}}{\dot{A_2}\nu}+1),\label{EOS}\\
\Omega_2&=&\frac{1}{1+\nu{A_2}(1-\Omega_{20})} ,\\
\Omega_x&=&1-\Omega_2,\\
q_2&=&\frac{1}{2}\left[{-\left(\frac{\dot{\nu}}{A_2\dot{A_2}}+\frac{\nu}{A_2^2}\right)
\frac{1-\Omega_{20}}{1+\nu{A_2}(1-\Omega_{20})}}+1\right],\label{q2}
\end{eqnarray}
where $\Omega_{20}$ is current value of matter density parameter
$\Omega_2$. If the function $\nu$ is given, the evolutions of all
cosmic observable parameters in (\ref{EOS})-(\ref{q2}) are
determined uniquely.

\section{Solutions to $\Lambda$CDM universe derived from global brane}
The Friedmann equations in four-dimensional $\Lambda$CDM universe
are
\begin{equation}
H^2+\frac{k}{a^2}=\frac{\kappa^2_{(4)}}{3}\rho+\frac{\Lambda}{3},\label{lcdm}
\end{equation}
\begin{equation}
\frac{\ddot{a}}{a}=-\frac{\kappa^2_{(4)}}{6}(\rho+3p)+\frac{\Lambda}{3},\label{lcdm2}
\end{equation}
where there is only the matter i.e $\rho=\rho_m$ and $p=p_m=0$.
Meanwhile, the equation of state on the cosmological constant
$\Lambda$ is $w_{\Lambda}=-1$. In order to get the LCDM universe
from the global brane, from Eq. (\ref{EOS}), we find when
\begin{equation}
p_x=-\rho_x,\label{-1}
\end{equation}
the property on the brane tends to $\Lambda$CDM universe. And, we
can find ${\kappa^2_{(4)}}={\kappa^2_{(5)}}/{(2y_2)}$. Since
$\kappa^2_{(5)}=M^{-3}_{(5)}$ and $\kappa^2_{(4)}=M^{-2}_{(4)}$, the
relation of the four dimensional Planck mass is expressed with five
dimensional Planck mass as
\begin{equation}
M^2_{(4)}=2y_2M^3_{(5)}.
\end{equation}
Therefore, the four dimensional Planck mass is relevant to five
dimensional Planck mass and the position of brane.

From Eq. (\ref{-1}), on the second brane i.e. $y=y_2\neq0$ we have
\begin{equation}
\frac{A_2\dot{\nu}}{\dot{A}_2\nu}=2.
\end{equation}
So, the relation of $\nu$ and $A_2$ is
\begin{equation}
\nu=CA_2^2,
\end{equation}
where $C$ is a integral constant. We can eliminate the arbitrary
function $\nu$ in Eq. (\ref{A}). Therefore, the (\ref{A}) is written
as
\begin{equation}
A_2^{2}=\left( \mu ^{2}+k\right)
y_2^{2}-2{CA_2^2}\mid{y_2}\mid+\frac{C^2A_2^4+K}{\mu ^{2}+k}.
\end{equation}
And this equation is rewritten as
\begin{equation}
\left[A^2_2-\frac{1}{2C^2}(2C|y_2|+1)(\mu^2+k)\right]^2
=\frac{1}{C^4}(|y_2|+\frac{1}{4})(\mu^2+k)^2-\frac{K}{C^2}.
\end{equation}
Therefore, the solutions of this equation are
\begin{equation}
A^2_2={\frac{1}{2C^2}(2C|y_2|+1)(\mu^2+k)}\pm\sqrt{\frac{1}{C^4}(|y_2|+\frac{1}{4})(\mu^2+k)^2-\frac{K}{C^2}},\label{A1}
\end{equation}
where for $A^2_2\geq{\frac{1}{2C^2}(2C|y_2|+1)(\mu^2+k)}$, ``$+$"
sign is taken; while for
$A^2_2\leq{\frac{1}{2C^2}(2C|y_2|+1)(\mu^2+k)}$, we choose ``$-$"
sign. This scale factor gives a clearer geometrical description
about the $5D$ spacetime. For $K=0$, we can find $I_1=I_2=I_3=0$ and
$5D$ bulk is a 5-dimensional flat spacetime. So, the scale factor in
the $5D$ flat spacetime is
\begin{equation}
A^2_2=\frac{1}{C^2}\left(C|y_2|+\frac{1}{2}\pm\sqrt{|y_2|+\frac{1}{4}}\right)(\mu^2+k).\label{A2}
\end{equation}
For a flat $3D$ space i.e. $k=0$, in the $5D$ flat spacetime, Eq.
(\ref{A2}) is simplified as
\begin{equation}
A^2_2=\frac{1}{C^2}\left(C|y_2|+\frac{1}{2}\pm\sqrt{|y_2|+\frac{1}{4}}\right)\mu^2.\label{0A2}
\end{equation}
Using the red-shift relation
\begin{equation}
A_2=\frac{A_{20}}{1+z},
\end{equation}
from the Eq.(\ref{0A2}), we have
\begin{equation}
\frac{1}{C^2}\left(C|y_2|+\frac{1}{2}\pm\sqrt{|y_2|+\frac{1}{4}}\right)\mu^2=\frac{A^2_{20}}{(1+z)^2}.
\end{equation}
So, for different $y_2$, the $A_{20}$ is different. In other words,
we obtain different $A_{20}$ in different brane. Therefore, if
considering when $z=0$, $A_{20}=1$ on different brane, the redshift
$z$ should be redefined on different brane.

In fact, these two solutions all can induce the $\Lambda$CDM
universe on $y\neq0$ brane. $C$ is an arbitrary constant.
Substituting $\nu=CA_2^2$ into Eq. (\ref{q2}), the deceleration
parameters is rewritten as
\begin{equation}
q_2=\frac{1}{2}\left[-\frac{3C(1-\Omega_{20})}{1+CA_2(1-\Omega_{20})}+1\right].
\end{equation}
The present value of deceleration parameters $q_{2}$ is
\begin{equation}
q_{20}=\frac{1}{2}\left[-\frac{3C(1-\Omega_{20})}{1+C(1-\Omega_{20})}+1\right].
\end{equation}
Our universe is accelerating, so the deceleration parameter is
$q_{20}<0$. Therefore, the range of $C$ is $C>1/(2-2\Omega_{20})$ or
$C<-1/(1-\Omega_{20})$. Adopting $q_{20}=-0.5$ and
$\Omega_{20}=0.3$, we have $C=20/7\approx3$.

If we utilize the result $\nu=CA^2_2$ on $y=0$ brane i.e
$\nu=CA^2_1$, the density on $y=0$ brane will be
$\rho_1=6C/\kappa^2_{(5)}$ and $p_1=-6C/\kappa^2_{(5)}$. Therefore,
the $y=0$ brane is dominated by abnormal matter with $w=-1$.
\section{Conclusions}
In this paper, the exact global solutions of brane universes are
discussed. The solutions contain two arbitrary functions $\mu$ and
$\nu$ of time $t$. On the brane, the Friedmann equations are
modified by the extra term with $\nu$. Therefore, the arbitrary
function $\nu$ will influence the evolution of our universe. Then we
find dimensionless density parameters and deceleration parameters
are relate to the arbitrary function $\nu$. $\Lambda$CDM universe is
the most simple model and not ruled out by present astronomical
observation. But we do not know where it comes from. Now, in order
to derive $\Lambda$CDM universe from the $5D$ spacetime, the
arbitrary function $\nu$ is eliminate in the scale factor of global
brane. So the more exact solutions of the global brane are obtained.
Then a clear $5D$ manifold is presented.

\section*{Acknowledgments}
This work was supported by NSF (10573003), NSF (10647110), NSF
(10703001), NBRP (2003CB716300) of P. R. China and DUT 893321.

\end{document}